\documentclass[10pt]{iopart}
\pdfoutput=1
\usepackage{graphicx}
\usepackage{amssymb}
\usepackage{xcolor}

\usepackage{iopams} 
\usepackage{listings}
\usepackage{float}
\restylefloat{table}

\eqnobysec
\newcommand{\beq}{\begin{equation}}
\newcommand{\beqa}{\begin{eqnarray}}
\newcommand{\eeq}{\end{equation}}
\newcommand{\eeqa}{\end{eqnarray}}
\renewcommand{\max}{{\rm max}}
\newcommand{\prob}{\mathop{\rm Prob}\nolimits}
\newcommand{\lap}[1]{\mathrel{\mathop{\cal L}\limits_{#1}^{}}}
\newcommand{\zeq}[1]{\mathrel{\mathop{=}\limits_{#1}^{}}}
\newcommand{\zapprox}[1]{\mathrel{\mathop{\approx}\limits_{#1}^{}}}
\newcommand{\num}{_{|{\rm num}}}
\newcommand{\T}{{\bf T}}
\newcommand{\comport}[3]{\mathrel{\mathop{#1}\limits_{#2}^{#3}}}

\begin{document}

\title[Two-time correlation function and occupation time]{Two-time correlation and occupation time for the Brownian bridge and tied-down renewal processes}
\date{\today}
\author{Claude Godr\`eche}
\address{
Institut de Physique Th\'eorique, Universit\'e Paris-Saclay, CEA and CNRS,
91191 Gif-sur-Yvette, France}\smallskip

\begin{abstract}
Tied-down renewal processes are generalisations of the Brownian bridge, where an event (or a zero crossing) occurs both at the origin of time and at the final observation time $t$.
We give an analytical derivation of the two-time correlation function for such processes in the Laplace space of all temporal variables.
This yields the exact asymptotic expression of the correlation in the Porod regime of short separations between the two times
and in the persistence regime of large separations.
We also investigate other quantities, such as the backward and forward recurrence times, as well as the occupation time of the process.
The latter has a broad distribution which is determined exactly.
Physical implications of these results for the Poland Scheraga and related models
are given.
These results also give exact answers to questions posed in the past in the context of stochastically evolving surfaces.

\end{abstract}

\maketitle
\section{Introduction}

Tied-down renewal processes with power-law distributions of intervals are generalisations of the Brownian bridge, where an event (or a zero crossing) occurs both at the origin of time and at the final observation time $t$~\cite{wendel,cgwendel}.
The Brownian bridge is itself the continuum limit of the tied-down simple random walk, starting and ending at the origin.
The present work is a sequel of our previous study~\cite{cgwendel} mainly devoted to the statistics of the longest interval of tied-down renewal processes.
Here we investigate further quantities of interest such as the two-time correlation function, the backward and forward recurrence times and the occupation time of the process.

The present study parallels that done in~\cite{gl2001}, where the statistics of these quantities were investigated in the unconstrained case (i.e., without the constraint of having an event at the observation time $t$)%
\footnote{The statistics of the longest interval for unconstrained renewal processes was investigated in~\cite{gms2015}.
},
then these results were used to give analytical insight in some simplified physical models.
The results found here for tied-down renewal processes provide analytical expressions of the pair correlation function in the Porod and persistence regimes and of the distribution of the magnetisation for the Poland Scheraga~\cite{poland} and related models~\cite{bar1,bar2}.
They also give exact answers to questions posed in the past in the context of stochastically evolving surfaces~\cite{das1,das2}.

This paper illustrates the importance of a systematic study of renewal processes given their ubiquity and potential applications in statistical physics.

We shall rely on~\cite{cgwendel} for some background knowledge, in order to keep the present paper short and avoid repeating the material contained in this reference.
Nevertheless, we shall start, in section~\ref{sec:defs}, by giving a brief reminder of the most important definitions needed in the subsequent sections~\ref{sec:excess}-\ref{sec:occup}, devoted respectively to the study of the statistics of the forward and backward recurrence times,
the number of renewals between two times, the two-time correlation function and finally the occupation time of the process.
Section~\ref{sec:discussion} gives applications of the present study to simple equilibrium or nonequilibrium physical systems.
Details of some derivations are relegated to appendices.

\section{Definitions}
\label{sec:defs}

\subsection{Renewal processes in general}
\label{sec:renewal}

We remind the definitions and notations used for renewal processes, following~\cite{gl2001}.
Events (or renewals) occur at the random epochs of
time $t_{1},t_{2},\ldots$, from some time origin $t=0$. 
These events are for instance the zero crossings of some stochastic process.
We take the origin of time on a zero crossing. 
When the intervals of time between events, 
$\tau_{1}=t_{1},\tau_{2}=t_{2}-t_{1},\ldots $, are independent and identically
distributed random variables with common density $\rho (\tau)$, the process thus
formed is a \textit{renewal process}~\cite{cox,feller}. 

The probability $p_0(t)$ that no event occurred up to time $t$
is simply given by the tail probability: 
\beq
p_{0}(t)=\prob(\tau_1 >t)=\int_{t}^{\infty }{\rm d}\tau\,\rho (\tau). 
\label{tail}
\eeq
The density $\rho(\tau)$ can be either a narrow distribution with all
moments finite, in which case the decay of $p_{0}(t)$, as $t\rightarrow \infty $,
is faster than any power law, or a distribution characterised by a
power-law fall-off with index $\theta>0 $ 
\beq\label{eq:pers}
p_0(t)=\int_{t}^{\infty }{\rm d}\tau\,\rho (\tau)
\approx \left( \frac{\tau_{0}}{t}\right) ^{\theta }, 
\eeq
where $\tau_{0}$ is a microscopic time scale. 
If $\theta <1$ all moments of 
$\rho(\tau) $ are divergent, if $1<\theta <2$, the first moment 
$\left\langle \tau\right\rangle $ is finite but higher moments are divergent, and so on. 
In Laplace space, where $s$ is conjugate to $\tau$, for a narrow distribution we have 
\begin{equation}
\fl \lap{\tau}\rho (\tau)=\hat{\rho}(s)=\int_0^\infty{\rm d}\tau\, \e^{-s \tau}\rho(\tau)
\zeq{ s\rightarrow 0}
1-\left\langle 
\tau\right\rangle s+\frac{1}{2}
\left\langle \tau^{2}\right\rangle s^{2}+\cdots \label{ro_narrow}
\end{equation}
For a broad distribution, (\ref{eq:pers}) yields 
\begin{equation}
\hat{\rho}(s)\zapprox{s\rightarrow 0}\left\{ 
\begin{array}{ll}
1-a\,s^{\theta } & (\theta <1) \\ 
1-\left\langle \tau\right\rangle s+a\,s^{\theta } & (1<\theta <2),
\end{array}
\right. \qquad \label{ro_broad}
\end{equation}
and so on, where 
\beq\label{eq:a}
a=|\Gamma (1-\theta )|\tau_{0}^{\theta }.
\eeq
From now on, unless otherwise stated, we shall only consider the case $0<\theta<1$.

The quantities naturally associated to a renewal process~\cite{gl2001,cox,feller} are the following.
The number of events which occurred between $0$ and $t$ (without counting the event at the origin), i.e., the largest $n$ such that $t_n\le t$, is a random variable denoted by $N_t$.
The time of occurrence of the last event before $t$, that is of the 
$N_t-$th event, is therefore the sum of a random number of random variables
\beq\label{eq:tN}
t_{N_t}=\tau_{1}+\cdots +\tau_{N_t}. 
\eeq
The backward recurrence time $A_{t}$ 
is defined as the length of time
measured backwards from $t$ to the last event before $t$, i.e., 
\beq
A_{t}=t-t_{N_t}.
\eeq
It is therefore the age of the current, unfinished, interval at time $t$.
Finally the forward recurrence time (or excess time) $E_{t}$ is the time
interval between $t$ and the next event 
\beq\label{eq:Edef}
E_{t}=t_{N_t+1}-t. 
\eeq
We have the simple relation $A_t+E_t=t_{N_t+1}-t_{N_t}=\tau_{N_t+1}$.
The statistics of these quantities is investigated in detail in~\cite{gl2001}.

\subsection{Tied-down renewal processes}
\label{sec:tied-down-renew}

A tied-down renewal process is defined by
the condition $\{t_{N_t}=t\}$, or equivalently by the condition $\{A_t=0\}$, which both express that the $N_t-$th event occurred at time $t$.
This process generalises the Brownian bridge~\cite{cgwendel,wendel}.

The joint density associated to the realisation $\{\ell_1,\dots,\ell_n\}$ of the sequence of $N_t=n$ intervals $\{\tau_1,\dots,\tau_n\}$, 
conditioned by $\{t_{N_t}=t\}$, is~\cite{cgwendel}
\beq\label{eq:fstar}
f^{\star}(t,\ell_1,\dots,\ell_n,n)
=\frac{\rho(\ell_1)\dots\rho(\ell_n)\delta\left(\sum_{i=1}^{n}\ell_i-t\right)}{U(t)},
\eeq
where the denominator is obtained from the numerator by integration on the $\ell_i$ and summation on $n$,
\beqa\label{eq:ftNdef+}
U(t)
&=&\sum_{n\ge0}\int_{0}^{\infty}{\rm d}\ell_1\dots{\rm d}\ell_n\,\rho(\ell_1)\dots\rho(\ell_n)\delta\Big(\sum_{i=1}^n\ell_i-t\Big).
\eeqa
This quantity
is the edge value of the probability density of $t_{N_t}$ at its maximal value $t_{N_t}=t$~\cite{cgwendel}.
In Laplace space with respect to $t$, we have
\beq\label{eq:lapUs}
\hat U(s)=\lap{t}U(t)=\sum_{n\ge0}\hat\rho(s)^n=\frac{1}{1-\hat\rho(s)}.
\eeq
The right side behaves, when $s$ is small, as $s^{-\theta}/a$.
Thus, at long times, we finally obtain, using~(\ref{eq:a}),
\beq\label{eq:denom}
U(t)\approx 
\frac{\sin \pi\theta}{\pi}\frac{t^{\theta-1}}{\tau_0^\theta }.
\eeq

Knowing the expression~(\ref{eq:fstar}) of the conditional density allows to compute the conditional average of any observable $O$, as
\beq
\langle O\rangle=\sum_{n\ge0}\int_{0}^{\infty}{\rm d}\ell_1\dots{\rm d}\ell_n\,f^{\star}(t,\ell_1,\dots,\ell_n,n)\,O.
\eeq
The method used in the next sections consists in computing separately the numerator of this expression, denoted by $\langle O\rangle\num$, then divide by the denominator, $U(t)$.

\section{Forward and backward recurrence times for the tied-down renewal process}
\label{sec:excess}

\begin{figure}[h]
\begin{center}
\includegraphics[angle=0,width=1\linewidth]{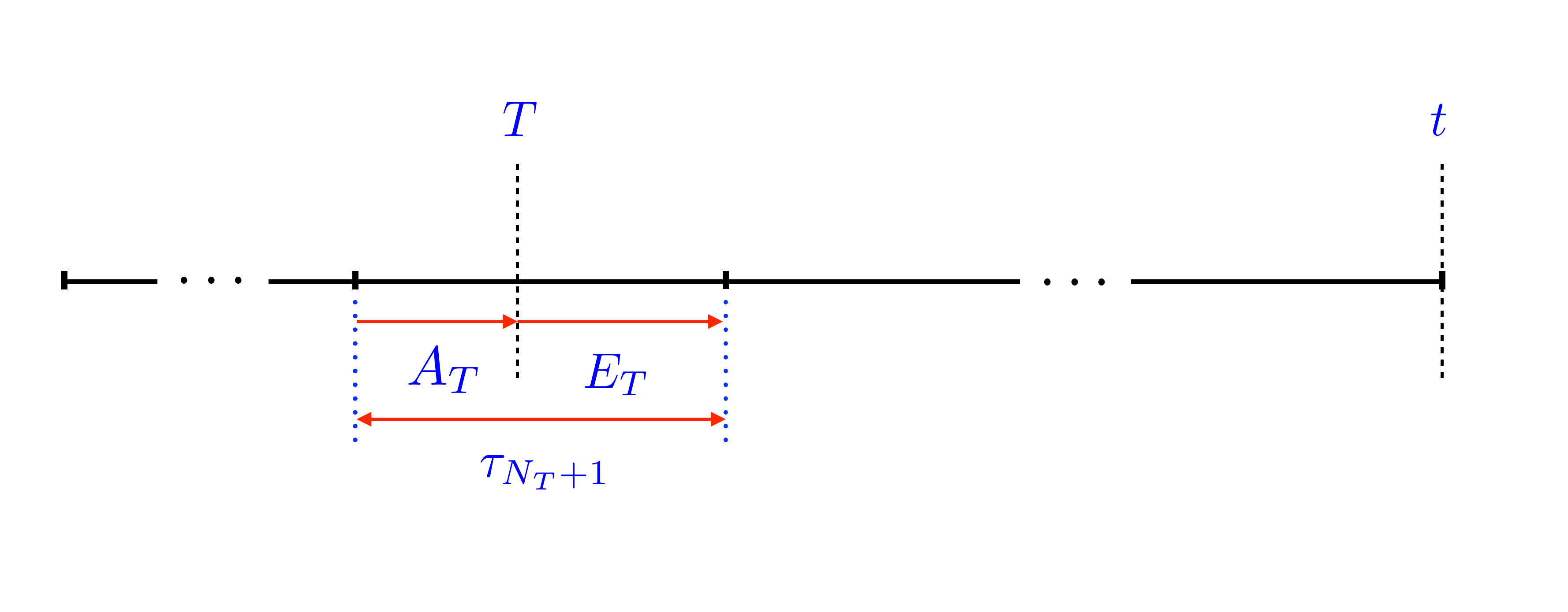}
\caption{The excess time (or forward recurrence time) $E_T$ with respect to $T$ 
is the distance between $T$ and the next event.
The age of the last interval before $T$ (or backward recurrence time) $A_T$ is the distance between $t_{N_t}$ and $T$.
The interval $\tau_{N_T+1}$ straddles $T$.
\label{fig:figEA}}
\end{center}
\end{figure}

Consider the situation depicted in figure~\ref{fig:figEA}.
The number of intervals between $0$ and the intermediate time $0<T<t$ is denoted by $N_T$.
This is also the number of points between 0 and $T$ (without counting the point at the origin).
Instead of~(\ref{eq:tN}) we have now
\beq\label{eq:tNT}
t_{N_T}=\tau_{1}+\cdots +\tau_{N_T}. 
\eeq
The excess time (or forward recurrence time) with respect to $T$ is, as in~(\ref{eq:Edef}),
the time interval between $T$ and the next event 
\beq
E_T=t_{N_{T}+1}-T,
\eeq
where $t_{N_{T}+1}=t_{N_{T}}+\tau_{N_{T}+1}$.
Its density is
\beq
f_E(t,T,y)=\langle \delta(y-E_T)\rangle=\frac{{\rm d}}{{\rm d} y}\prob(E_T<y|t_{N_t=t}).
\eeq
In Laplace space, where $s,u,v$ are conjugate to the temporal variables $t,T,y$, we find for the numerator (see~\ref{app:fElap})%
\footnote{When no ambiguity arises, we drop the time dependence of the random
variable if the latter is itself in subscript.},
\beq\label{eq:fElap}
\fl\hat f_E(s,u,v){\num}=
\lap{t,T,y}f_E(t,T,y){\num}=\hat U(s)\hat U(s+u)\frac{\hat\rho(s+v)-\hat\rho(s+u)}{u-v}.
\eeq
This expression will be used in the next section.

The density $f_E(t,T,y)$ is well normalised, as can be seen by noting that
\beqa
\fl\hat f_E(s,u,0){\num}=\hat U(s)\hat U(s+u)\frac{\hat\rho(s)-\hat\rho(s+u)}{u}
&=&\frac{\hat U(s)-\hat U(s+u)}{u}
\nonumber\\
&=&\lap{t,T}U(t)\Theta(t-T),
\label{eq:PNN}
\eeqa
where $\Theta$ is the Heaviside function.
So, dividing by $U(t)$,
\beq
\int_0^\infty{\rm d} y\,f_E(t,T,y)=\Theta(t-T).
\eeq

Similarly, one finds for the forward recurrence time $A_T=T-t_{N_T}$, also named the age of the last interval before $T$ (see~\ref{app:fAlap}),
\beq\label{eq:fAlap}
\fl\hat f_A(s,u,v){\num}=
\lap{t,T,y}f_A(t,T,y){\num}=\hat U(s)\hat U(s+u)\frac{\hat\rho(s)-\hat\rho(s+u+v)}{u+v}.
\eeq

\subsubsection*{Remark.} 

The limit $t\to\infty$ ($s\to0$) corresponds to the unconstrained case.
In this limit (\ref{eq:fElap}) reads
\beq
\fl\hat f_E(s\to0,u,v){\num}=
\hat U(s\to0)\hat U(u)\frac{\hat\rho(v)-\hat\rho(u)}{u-v},
\eeq
which, after Laplace inversion with respect to $s$ and division by $U(t)$, yields, as it should, the expression found for the unconstrained case (see equation (6.2) in~\cite{gl2001}) up to the change of $u$ to $s$ and $v$ to $u$.
The same holds for $f_A$.

\section{Number of renewals between two arbitrary times}

\begin{figure}[h]
\begin{center}
\includegraphics[angle=0,width=1\linewidth]{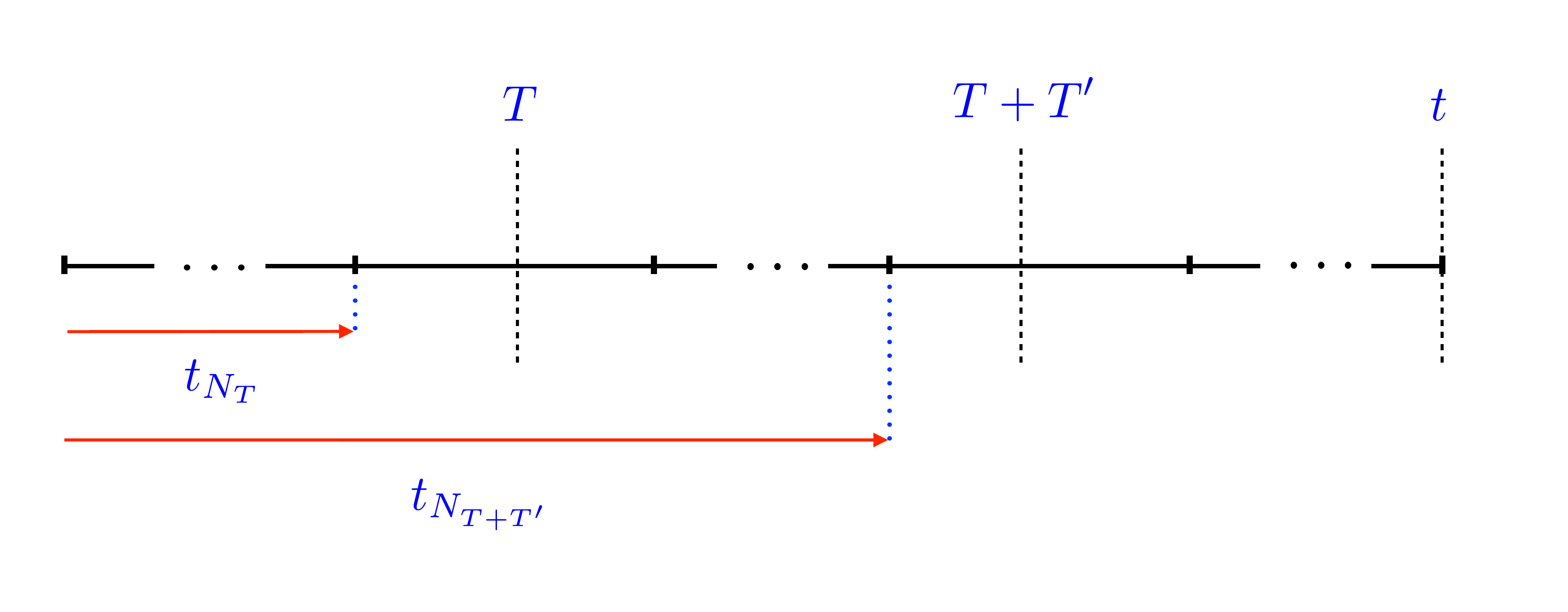}
\caption{There are $N_{T}$ events up to time $T$ and $N(T,T+T')$ events between $T$ and $T+T'$.
\label{fig:2timeB}}
\end{center}
\end{figure}

Consider the number of events $N(T,T+T')=N_{T+T'}-N_{T}$ occurring between the two times $T$ and $T+T'$ (see figure~\ref{fig:2timeB}).
We denote the probability distribution of this random variable by
\beq
p_{m}(t,T,T+T')=\prob(N(T,T+T')=m|t_{N_t}=t).
\eeq
For $m=0$ we have
\beq\label{eq:p0tTT}
p_0(t,T,T+T')=\prob(E_T>T')=\int_{T'}^\infty
{\rm d}{y}\,f_E(t,T,y).
\eeq
In Laplace space with respect to the three temporal variables $t,T,T'$, we find, for $m\ge1$ (see~\ref{app:pmtTT}),
\beqa\label{eq:pm}
\hat p_m(s,u,v)\num
&=&\lap{t,T,T'}p_{m}(t,T,T+T')\num
\nonumber\\
&=&\hat f_E(s,u,v)\num\frac{\hat\rho(s)-\hat\rho(s+v)}{v}\hat\rho(s+v)^{m-1},
\eeqa
and, for $m=0$,
\beqa\label{eq:p0lap}
\hat p_0(s,u,v)\num
&=&\lap{t,T,T'}p_0(t,T,T+T')\num
\nonumber\\
&=&
\frac{1}{v}\left(\hat f_E(s,u,0){\num}-\hat f_E(s,u,v){\num}\right),
\eeqa
which is a simple consequence of~(\ref{eq:p0tTT}) (an alternative proof is given in~\ref{app:pmtTT}).
The scaling behaviour of $p_0(t,T,T+T')$ in the temporal domain is analysed in the next section.

The normalisation of the distribution $p_m$ can be checked by computing the sum
\beq\label{eq:norma}
\fl\sum_{m\ge0}\hat p_m(s,u,v)\num
=\frac{1}{v}\left(\frac{\hat U(s)-\hat U(s+u)}{u}-\frac{\hat U(s+v)-\hat U(s+u)}{u-v}\right),
\eeq
the inverse Laplace transform of which is
\beqa
\sum_{m\ge0}p_m(t,T,T+T')\num
&=&U(t)\Theta(t-T)\Theta(t-T-T'),
\eeqa
as can be easily checked (see~(\ref{eq:PNN})).
So, finally, after division by $U(t)$,
\beq
\sum_{m\ge0}p_m(t,T,T+T')
=\Theta(t-T)\Theta(t-T-T').
\eeq

\section{Two-time correlation function}
\label{sec:2time}

\begin{figure}[h]
\begin{center}
\includegraphics[angle=0,width=1\linewidth]{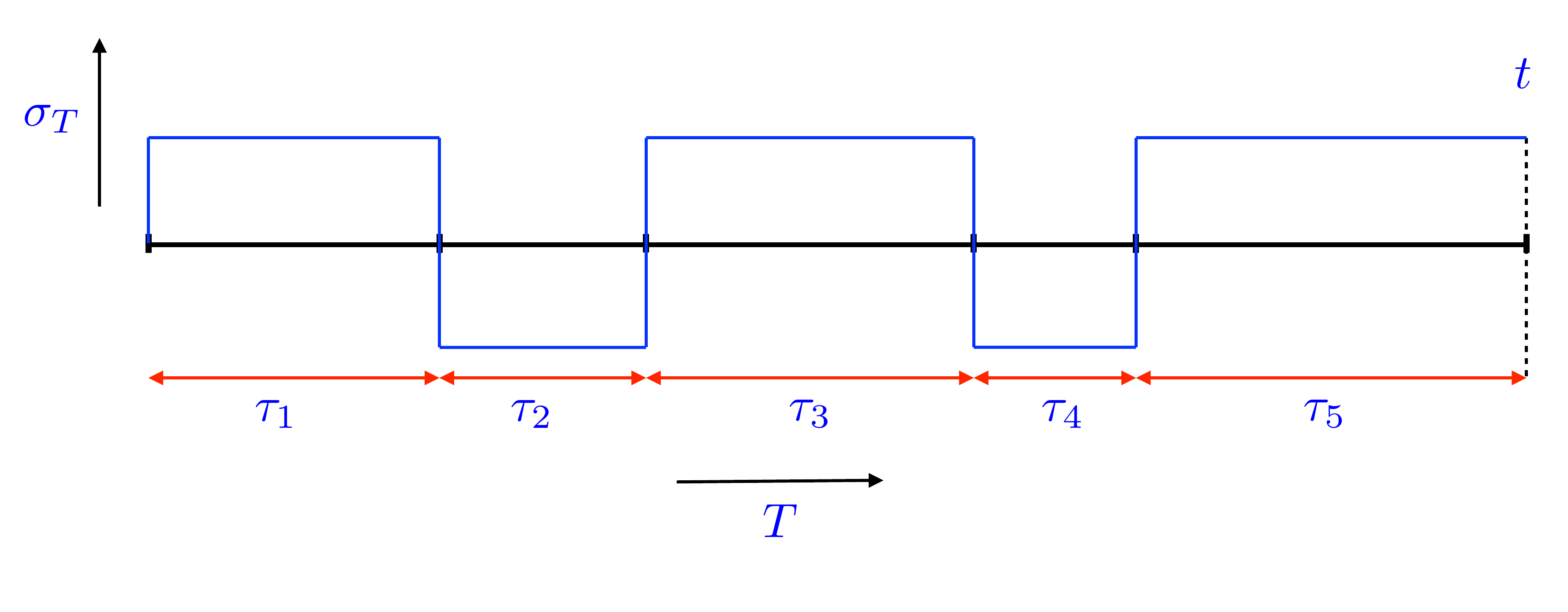}
\caption{The $\sigma_T=\pm1$ process where $T$ is the running time.
In this example the initial condition is $\sigma_0=+1$.
\label{fig:figsigma}}
\end{center}
\end{figure}
In the present section and the next one, we consider the random variable $\sigma_T=\pm1$, where $T$ is the running time, linked to the renewal process as follows.
Let $\sigma_T=1$, for all the duration of the first interval, then $\sigma_T=-1$, during the second interval, and so on, with alternating values, as depicted in figure~\ref{fig:figsigma}.
This process can be thought of as the sign of the position of a one-dimensional underlying motion (such as Brownian motion if $\theta=1/2$).
In other words the random variable $\sigma_T$ is the sign of the successive excursions (positive if the motion is on the right side of the origin, negative otherwise).
We can also interpret $\sigma_T$ as a spin variable (depending on time $T$) and the intervals $\tau_1,\tau_2,\dots$ as the intervals between two flips~\cite{gl2001,balda}.
In the present case, the tied-down constraint imposes that the sum of these intervals is given.

\subsection{Analytic expression of the two-time correlation function in Laplace space}
We want to compute the correlation
\beqa\label{eq:correlation}
\fl C(t,T,T+T')=\langle\sigma_{T}\sigma_{T+T'}\rangle=\langle (-1)^{N(T,T+T')}\rangle
=\sum_{m\ge0}(-1)^m
p_m(t,T,T+T'),
\eeqa
between the two times $T$ and $T+T'$.
Using the expressions~(\ref{eq:pm}) and~(\ref{eq:p0lap}) above, we obtain, in Laplace space,
\beq\label{eq:resC}
\fl\hat C(s,u,v)\num=\frac{1}{v}\left(\hat f_E(s,u,0)\num-\hat f_E(s,u,v)\num \frac{1+\hat\rho(s)}{1+\hat\rho(s+v)}\right).
\eeq

Taking successively the limits $t\to\infty$, $T\to0$ and $T'\to0$ allows to check the coherence of the formalism.
\begin{enumerate}
\item The limit $t\to\infty$ ($s\to0$) corresponds to the unconstrained case, as already noted above.
We find, after Laplace inversion with respect to $s$ and division by $U(t)$,
\beq
\hat C(s\to0,u,v)=
\frac{1}{v}\left(\frac{1}{u}-\hat U(u)\frac{\hat\rho(v)-\hat\rho(u)}{u-v}\frac{2}{1+\hat\rho(v)}\right).
\eeq
Changing $u$ to $s$, and $v$ to $u$ in this expression yields equation (9.1) of~\cite{gl2001}, which is the Laplace transform of the two-time correlation in the unconstrained case.
\item In order to take the limit $T\to0$, we multiply~(\ref{eq:resC}) by $u$ then take the limit $u\to\infty$.
This yields
\beq
\fl\lim_{u\to\infty}u\,\hat C(s,u,v)\num=\hat U(s)
\frac{1}{v}\frac{\hat\rho(s)-\hat\rho(s+v)}{1+\hat\rho(s+v)}\equiv
\lap{t,T'}\langle(-1)^{N_{T'}}\rangle\num,
\eeq
which, after Laplace inversion and division by $U(t)$, yields the correlation function $\langle \sigma_0\sigma_{T'}\rangle$.
\item
Finally the limit $T'\to0$ is obtained by computing
\beq
\fl\lim_{v\to\infty}v\,\hat C(s,u,v)\num=\hat f_E(s,u,0)\num,
\eeq
which, after Laplace inversion and division by $U(t)$, yields unity as expected.
\end{enumerate}

\subsection{Asymptotic analysis in the Porod regime}
At large and comparable times $t,T,T'$, we have $s\sim u\sim v\ll1$, hence 
\beq
\hat C(s,u,v)\num\approx \hat p_0(s,u,v)\num,
\eeq
which is given by~(\ref{eq:p0lap}).
In the present context, $p_0(t,T,T+T')$ is the probability that the spin $\sigma_T$ did not flip between $T$ and $T+T'$, or two-time persistence probability.
The inversion of the first term in the right side of~(\ref{eq:p0lap}) yields $U(t)\Theta(t-T)$, which, after division by $U(t)$ yields 1 (since $t>T$).
Let us analyse the second term in the right side of~(\ref{eq:p0lap}).
In the regime $s\sim u\sim v$ we have
\beq\label{eq:res1}
\frac{1}{v}\hat f_E(s,u,v)\num\approx
\frac{1}{v}\frac{1}{a s^\theta(s+u)^\theta}\frac{(s+u)^\theta-(s+v)^\theta}{u-v}.
\eeq
Let us moreover focus on the regime of short separations between $T$ and $T+T'$, i.e., where $1\ll T'\ll T\sim t$.
In this regime we expect the following scaling form for the two-time correlation function
\beq\label{eq:ansatz}
C(t,T,T+T')\approx 1-\left(\frac{T'}{t}\right)^{1-\theta}g\left(\frac{T}{t}\right),
\eeq
where the scaling function $g(\cdot)$ is to be determined.
In the regime of interest ($s\sim u\ll v$), (\ref{eq:res1}) simplifies into
\beq
\frac{1}{v}\hat f_E(s,u,v)\num\approx
\frac{v^{\theta-2}}{a s^\theta(s+u)^\theta}.
\eeq
Laplace inverting the right side of this equation with respect to $v$ yields
\beq
\frac{T'^{1-\theta}}{a\Gamma(2-\theta)s^\theta(s+u)^\theta},
\eeq
which, after Laplace inversion with respect to $s$ and $u$ and division by $U(t)$, should be identified with the second term of~(\ref{eq:ansatz}), i.e.,
\beq
\frac{1}{U(t)}\frac{T'^{1-\theta}}{a\Gamma(2-\theta)}\lap{s,u}^{-1}\frac{1}{s^\theta(s+u)^\theta}
=\left(\frac{T'}{t}\right)^{1-\theta}g\left(\frac{T}{t}\right).
\eeq
We thus have the following identification to perform
\beq\label{eq:identif}
\lap{s,u}^{-1}\frac{1}{s^\theta(s+u)^\theta}
={\Gamma(1-\theta)\Gamma(2-\theta)}\frac{\sin\pi\theta}{\pi}t^{2\theta-2}g\left(\frac{T}{t}\right).
\eeq
We have, setting $x=T/t$,
\beqa
\lap{t,T}t^{2\theta-2}g\left(\frac{T}{t}\right)
&=&\int_0^1{\rm d} x\,g(x)\int_0^\infty{\rm d} t\, t^{2\theta-1}\e^{-t(s+ux)}
\nonumber\\
&=&\frac{\Gamma(2\theta)}{s^{2\theta}}\int_0^1{\rm d} x\,\frac{g(x)}{(1+ux/s)^{2\theta}}.
\eeqa
So, using~(\ref{eq:identif}), we finally obtain the equation determining the scaling function $g(x)$, 
\beq\label{eq:stiel}
\int_0^1{\rm d} x\,\frac{g(x)}{(1+\lambda x)^{2\theta}}
=\frac{\Gamma(\theta)}{\Gamma(2-\theta)\Gamma(2\theta)}
\frac{1}{(1+\lambda)^\theta},
\eeq
with the notation $\lambda=u/s$.
Noting that
\beq
\int_0^1{\rm d} x\,\frac{\left[x(1-x)\right]^{\theta-1}}{(1+\lambda x)^{2\theta}}
=\frac{\Gamma(\theta)^2}{\Gamma(2\theta)(1+\lambda)^\theta},
\eeq
we conclude that
\beq
g(x)=\frac{\sin\pi\theta}{\pi(1-\theta)}\left[x(1-x)\right]^{\theta-1},
\eeq
yielding the final result, in the regime of short separations between $T$ and $T+T'$ ($1\ll T'\ll T\sim t$),
\beqa\label{eq:fdpo}
C(t,T,T+T')
&\approx &
1-\frac{\sin\pi\theta}{\pi(1-\theta)}\left[\frac{T}{t}\left(1-\frac{T}{t}\right)\right]^{\theta-1}
\left(\frac{T'}{t}\right)^{1-\theta}
\nonumber\\
&\approx&1-\frac{\sin\pi\theta}{\pi(1-\theta)}\left(\frac{T't}{T(t-T)}\right)^{1-\theta}.
\eeqa
This expression is universal since it no longer depends on the microscopic scale $\tau_0$.
When $t\to\infty$ we recover the result, easily extracted from~\cite{gl2001}, for the unconstrained case in the same regime ($1\ll T'\ll T$), namely
\beqa
C(T,T+T')
\approx 
1-\frac{\sin\pi\theta}{\pi(1-\theta)}\left(\frac{T'}{T}\right)^{1-\theta}.
\eeqa

\subsubsection*{Remark.}
Let us apply this formalism to the case of an exponential distribution of intervals, $\rho(\tau)=\lambda\e^{-\lambda\tau}$ corresponding to a Poisson process for the renewal events.
We have, from~(\ref{eq:resC}),
\beq
\fl\hat C(s,u,v)\num=\frac{\lambda}{s(s+u)(s+v+2\lambda)}.
\eeq
By Laplace inversion we obtain
\beq
C(t,T,T+T')=\e^{-2\lambda T'}\Theta(t-T)\Theta(t-T-T'),
\eeq
where we used the fact that $U(t)=\lambda$.
The correlation is stationary, i.e., a function of $T'$ only, as expected.
For time differences $T'$ such that $\lambda T'$ is small, i.e., $T'\ll \lambda^{-1}$, this correlation reads $C(T')\approx 1-2\lambda T'$.
Interpreted in the spatial domain, this expression is the usual Porod law~\cite{bray}, where $\lambda$ is the density of defects (domain walls).

\subsection{Asymptotic analysis in the persistence regime}
Let us now focus on the regime of large separations between $T$ and $T+T'$, i.e., where $1\ll T\ll T'\sim t$.
In this regime we expect the following scaling form for the two-time correlation function
\beq\label{eq:ansatzP}
C(t,T,T+T')\approx p_0(t,T,T+T')\approx \left(\frac{T'}{T}\right)^{-\theta}
h\left(\frac{T'}{t}\right),
\eeq
where the scaling function $h(\cdot)$ is to be determined.
In the regime of interest ($s\sim v\ll u$), (\ref{eq:res1}) simplifies into
\beq
\frac{1}{v}\hat f_E(s,u,v)\num\approx
\frac{1}{v}\frac{1}{a s^\theta u^\theta}\frac{u^\theta-(s+v)^\theta}{u}.
\eeq
Therefore, 
\beqa\label{eq:p0lapP}
\hat p_0(s,u,v)\num
&=&
\frac{1}{v}\left(\hat f_E(s,u,0){\num}-\hat f_E(s,u,v){\num}\right)
\nonumber\\
&\approx&
\frac{1}{v}\frac{1}{a s^\theta u^\theta}\frac{(s+v)^\theta-s^\theta}{u}.
\eeqa

We proceed as above.
Laplace inverting the right side of this equation with respect to $u$ yields
\beq
\frac{T^{\theta}}{a\Gamma(1+\theta)}\frac{(s+v)^\theta-s^\theta}{v s^\theta},
\eeq
which, after Laplace inversion with respect to $s$ and $v$ and division by $U(t)$, should be identified with the second term of~(\ref{eq:ansatzP}), i.e.,
\beq
\frac{1}{U(t)}\frac{T^{\theta}}{a\Gamma(1+\theta)}\lap{s,v}^{-1}\frac{(s+v)^\theta-s^\theta}{v s^\theta}
=\left(\frac{T'}{T}\right)^{-\theta}
h\left(\frac{T'}{t}\right).
\eeq
We thus have the following identification to perform
\beq\label{eq:identifP}
\lap{s,v}^{-1}\frac{(s+v)^\theta-s^\theta}{v s^\theta}
=\frac{\theta}{t}
\left(\frac{T'}{t}\right)^{-\theta}
h\left(\frac{T'}{t}\right).
\eeq
We have, setting $x=T'/t$,
\beqa
\lap{t,T'}\frac{\theta}{t}\left(\frac{T'}{t}\right)^{-\theta}
h\left(\frac{T'}{t}\right)
&=&\theta
\int_0^1{\rm d} x\,x^{-\theta}h(x)\int_0^\infty{\rm d} t\, \e^{-t(s+vx)}
\nonumber\\
&=&\frac{\theta}{s}\int_0^1{\rm d} x\,\frac{x^{-\theta}h(x)}{1+xv/s}.
\eeqa
So, using~(\ref{eq:identifP}), we finally obtain the equation determining the scaling function $h(x)$, 
\beq\label{eq:stielP}
\int_0^1{\rm d} x\,\frac{x^{-\theta}h(x)}{1+\lambda x}
=\frac{(1+\lambda)^{\theta}-1}{\lambda},
\eeq
with the notation $\lambda=v/s$.
Hence
\beq
h(x)=\frac{\sin\pi\theta}{\pi\theta}\left(1-x\right)^{\theta},
\eeq
yielding the final result, in the regime of large separations between $T$ and $T+T'$ ($1\ll T\ll T'\sim t$),
\beqa\label{eq:fdpoP}
\fl C(t,T,T+T')\approx p_0(t,T,T+T')
&\approx &\frac{\sin\pi\theta}{\pi\theta}
\left(\frac{T'}{T}\right)^{-\theta}
\left(1-\frac{T'}{t}\right)^{\theta}
\nonumber\\
&\approx&\frac{\sin\pi\theta}{\pi\theta}
\left(\frac{T't}{T(t-T')}\right)^{-\theta}.
\eeqa
This expression is universal since it no longer depends on the microscopic scale $\tau_0$.
When $t\to\infty$ we recover the result, easily extracted from~\cite{gl2001}, for the unconstrained case in the same regime ($1\ll T\ll T'$), namely
\beqa
C(T,T+T')\approx p_0(t,T,T+T')
\approx 
\frac{\sin\pi\theta}{\pi \theta}\left(\frac{T'}{T}\right)^{-\theta}.
\eeqa


%
\subsection{Brownian bridge}
For the Brownian bridge, the two-time correlation function~(\ref{eq:correlation}) has an explicit expression.
Let $t_1$ and $t_2$ be two arbitrary times and $x_{t_1}$ and $x_{t_2}$ the corresponding positions of the process.
Then
\beq
\langle\sigma_{t_1}\sigma_{t_2}\rangle=
\frac{2}{\pi}
\arcsin\frac{\langle x_{t_1}x_{t_2}\rangle}{\sqrt{\langle x_{t_1}^2\rangle\langle x_{t_2}^2\rangle}}.
\eeq
For the Brownian bridge between $0$ and $t$, the correlation of positions reads
\beq
\langle x_{t_1}x_{t_2}\rangle=\frac{(t-t_2)t_1}{t}.
\eeq
So
\beq
\langle\sigma_{t_1}\sigma_{t_2}\rangle=
\frac{2}{\pi}
\arcsin\sqrt{\frac{t_1(t-t_2)}{t_2(t-t_1)}}.
\eeq
In the present case, $t_1\equiv T$, $t_2\equiv T+T'$.
Hence the result
\beqa
C(t,T,T+T')
&=&\frac{2}{\pi}
\arcsin\sqrt{\frac{T(t-T-T')}{(T+T')(t-T)}}
\nonumber\\
&=&1-\frac{2}{\pi}\arccos\sqrt{\frac{T(t-T-T')}{(T+T')(t-T)}}.
\eeqa
In the regime of short separations between $T$ and $T+T'$ ($1\ll T'\ll T\sim t$), we obtain
\beq
C(t,T,T+T')\approx1-\frac{2}{\pi}\sqrt{\frac{T't}{T(t-T)}},
\eeq
which is~(\ref{eq:fdpo}) with $\theta=1/2$.
In the regime of large separations between $T$ and $T+T'$ ($1\ll T\ll T'\sim t$), we obtain
\beq
C(t,T,T+T')\approx p_0(t,T,T+T')\approx
\frac{2}{\pi}\sqrt{\frac{T(t-T')}{T't}},
\eeq
which is~(\ref{eq:fdpoP}) with $\theta=1/2$.

\section{Occupation time}
\label{sec:occup}

We turn to an investigation of the occupation time $\T_t$ spent by the process $\sigma_T$ in the $(+)$ state up to time $t$, namely
\beq
\T_t=\int_{0}^t {\rm d} T \frac{1+\sigma_{T}}{2}.
\eeq
We also consider the more symmetrical quantity
\beq\label{eq:defS}
S_t=\int_{0}^t {\rm d} T\, \sigma_{T}=2\T_t-t.
\eeq
We follow the line of thought of~\cite{gl2001}, which tackles the unconstrained case, in order to analyse these observables.
The expression of $\T_t$ depends both on the initial condition $\sigma_0$ and on the parity of the number of intervals $N_t$.
If $\sigma_0=+1$, then
\beqa
\T_t=\tau_1+\tau_3+\cdots+\tau_{2k+1}\qquad &({\rm if}\ N_t=2k+1)
\\
\T_t=\tau_1+\tau_3+\cdots+\tau_{2k-1}\qquad &({\rm if}\ N_t=2k).
\eeqa
The first line is illustrated in figure~\ref{fig:figsigma}.
If $\sigma_0=-1$, then
\beqa
\T_t=\tau_2+\tau_4+\cdots+\tau_{2k}\qquad &({\rm if}\ N_t=2k+1)
\\
\T_t=\tau_2+\tau_4+\cdots+\tau_{2k}\qquad &({\rm if}\ N_t=2k).
\eeqa

The probability density of $\T_t$ is
\beq
 f_{\T}(t,y)=\langle \delta(y-\T_t)\rangle=\sum_{k\ge0}f_{\T,N_t}(t,y,k).
\eeq
In Laplace space, with $s,u$ conjugate to the temporal variables $t,y$, we find, for the numerator, if $\sigma_0=+1$,
\beqa
\fl\lap{t,y}f_{\T,N_t}(t,y,k){\num}
=\lap{t}\langle\e^{-u\T_t}\rangle\num
= \hat f_{\T,N_t}(s,u,k){\num}
\\
= \left\{
\begin{array}{ll}
\hat\rho(s+u)^{k+1}\hat\rho(s)^k & (N_t=2k+1)\vspace{12pt}
\\ 
\hat\rho(s+u)^{k}\hat\rho(s)^k & (N_t=2k).
\end{array}
\right.
\eeqa
If $\sigma_0=-1$,
\beqa
\hat f_{\T,N_t}(s,u,k){\num}
\\
= \left\{
\begin{array}{ll}
\hat\rho(s+u)^{k}\hat\rho(s)^{k+1} & (N_t=2k+1)\vspace{12pt}
\\ 
\hat\rho(s+u)^{k}\hat\rho(s)^k & (N_t=2k),
\end{array}
\right.
\eeqa
Summing on $k$, and adding the two contributions corresponding to $\sigma_0=\pm1$ with equal weight $1/2$, we finally find
\beq
\hat f_{\T}(s,u)\num=
\frac{1}{1-\hat\rho(s)\hat\rho(s+u)}
\left(1+\frac{\hat\rho(s)+\hat\rho(s+u)}{2}\right).
\eeq

Let us analyse this expression in the regime, $1\ll t\sim y$, with $y/t=x$ fixed, i.e., such that $s\sim u\ll 1$, with $u/s=\lambda$ fixed.
This yields
\beq
\hat f_{\T}(s,u)\num\approx \frac{2}{	a\left(s^\theta+(s+u)^\theta\right)}.
\eeq
This expression should be identified to
\beq
\lap{t,y}U(t) f_{\T}(t,y)\comport{\approx}{}{}\lap{t,y}\frac{U(t)}{t} f_{t^{-1}\T}(t,x=y/t).
\eeq
In this regime, the limiting density $f_{t^{-1}\T}(t,x=y/t)$, denoted by $f_X(x)$, of the random variable 
\beq
X=\lim_{t\to\infty}t^{-1}\T_t,
\eeq
 no longer depends on time $t$.
So, we are left with the equation for the unknown density $f_{X}(x)$,
\beqa
\frac{2}{a\left(s^\theta+(s+u)^\theta\right)}
&=&\int_{0}^\infty{\rm d} x\, f_{X}(x)\int_0^\infty{\rm d} t\, \e^{-t(s+ux) }U(t)
\nonumber\\
&=&\int_0^1{\rm d} x\,f_{X}(x)\hat U(s+ux),
\eeqa
that is
\beq\label{eq:stielt}
\int_0^1{\rm d} x\,\frac{f_{X}(x)}{(1+\lambda x)^\theta}
=\left\langle \frac{1}{(1+\lambda X)^\theta}\right\rangle
=\frac{2}{1+(1+\lambda)^\theta},
\eeq
where the brackets correspond to averaging on the density $f_{X}(x)$ of $X$.
Let us note that $f_{X}(x)$ is well normalised, as can be seen by setting $\lambda=0$ in both sides of the equation.
A second remark is that the result obtained is universal since the microscopic scale $\tau_0$ is no longer present.

\begin{figure}[h]
\begin{center}
\includegraphics[angle=0,width=1\linewidth]{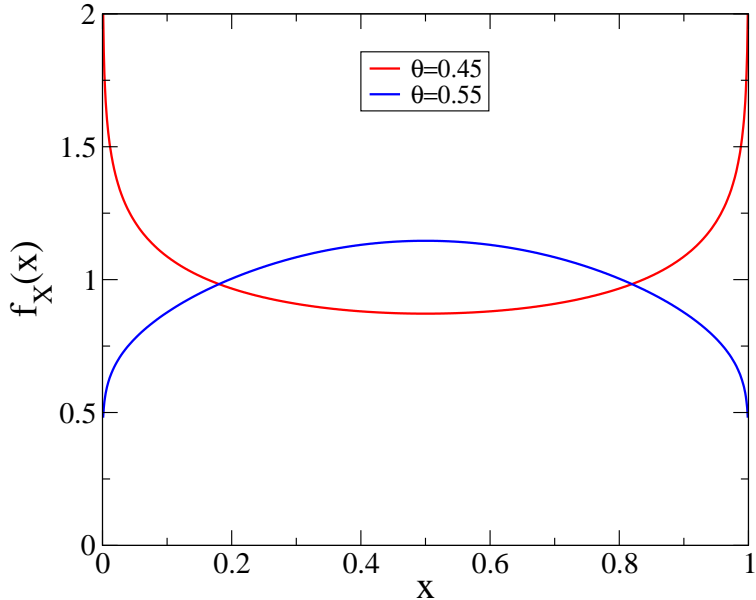}
\caption{Density of the scaled occupation time $X=\lim_{t\to\infty}t^{-1}\T_t$ for two values of the tail exponent $\theta$,
illustrating the change of concavity when crossing $\theta=1/2$ (Brownian bridge).
For this latter value  the distribution is uniform between $0$ and $1$.
The two values of $\theta$ in the figure were chosen not too far from $1/2$
because the further from this value, the more difficult is the numerical evaluation of the distribution of the occupation time.
\label{fig:occ}}
\end{center}
\end{figure}

For $\theta=1$ the left side of~(\ref{eq:stielt}) is the usual Stieltjes transform of $f_{X}(x)$, with solution $f_{X}(x)=\delta(x-1/2)$.
For $\theta=1/2$, the solution is $f_{X}(x)=1$.
We thus recover the well-known result, attributed to L\'evy, stating that the occupation time of the Brownian bridge is uniform on $(0,t)$.
More generally, the left side of~(\ref{eq:stielt}) is the generalised Stieltjes transform of index $\theta$ of $f_{X}(x)$\footnote{Note a first occurrence of the generalised Stieltjes transform in~(\ref{eq:stiel}).}.
A similar equation can be found in~\cite{yano1,yano}, in the context of the occupation time of Bessel bridges.
Let $F(x)=\int_0^x{\rm d} u\, f_{X}(u)$.
Then $F(x)$ is given by the fractional integral~\cite{yano}
\beq
F(x)=\int_0^x{\rm d} u\,(x-u)^{\theta-1}h(u),
\eeq
where
\beq
\fl h(u)=\frac{\sin\pi\theta}{\pi}\frac{2u^\theta}{u^{2\theta}+(1-u)^{2\theta}+2 u^\theta (1-u)^\theta\cos\pi\theta}\qquad (0<u<1).
\eeq
So the result is
\beq\label{eq:fX}
f_{X}(x)=\int_0^x{\rm d} u\,(x-u)^{\theta-1}h'(u).
\eeq
This distribution is U-shaped for $\theta<1/2$, and has its concavity inverted for $\theta>1/2$ (see figure~\ref{fig:occ}).
\textcolor{black}{It is universal with respect to the choice of distribution of intervals $\rho(\tau)$ as it only depends on the tail exponent $\theta$.}
For $x\to0$, the density $f_X(x)$ behaves as
\beq
f_X(x)\approx 
\frac{2\Gamma(1+\theta)}{\Gamma(1-\theta)\Gamma(2\theta)}
x^{2\theta-1}.
\eeq
For $\theta<1/2$ the density diverges at the origin, while for $\theta>1/2$ it vanishes.

Expanding the left and right sides of~(\ref{eq:stielt}) yields the moments of the distribution $f_{X}(x)$:
\beqa
\langle X\rangle=\frac{1}{2},\quad
\langle X^2\rangle=\frac{1}{2(1+\theta)},\quad
\langle X^3\rangle=\frac{2-\theta}{4(1+\theta)},
\nonumber\\
\langle X^4\rangle=\frac{3(2-\theta^2)}{2(1+\theta)(2+\theta)(3+\theta)},
\eeqa
and so on.
For $\theta=1/2$ one recovers the moments of the uniform distribution on $(0,1)$.

Coming back to the quantity $S_t$, we have, using~(\ref{eq:defS}),
\beq
\hat f_S(s,u)=\hat f_{\T}(s-u,2u),
\eeq
so
\beq
\fl\hat f_{S}(s,u)\num=
\frac{1}{1-\hat\rho(s-u)\hat\rho(s+u)}
\left(1+\frac{\hat\rho(s-u)+\hat\rho(s+u)}{2}\right).
\eeq
Here $\hat f_{S}(s,u)$ is the bilateral Laplace transform of $f_{S}(t,y)$ with respect to $y$ (see~\cite{gl2001}) and its usual Laplace transform with respect to $t$.
The scaled quantity $M_t=t^{-1}S_t$ has, when $t\to\infty$, the limiting density
\beq\label{eq:fM}
f_M(m)=\frac{1}{2}f_{X}\left(\frac{1+m}{2}\right),
\eeq
where $-1<m<1$, with vanishing odd moments and
\beq
\langle M^2\rangle=\frac{1-\theta}{1+\theta},\quad
\langle M^4\rangle=\frac{(1-\theta)(6-5\theta)}{(2+\theta)(3+\theta)}, 
\eeq
and so on.
Considering time as a distance, $M_t$ has the interpretation of the magnetisation of a simple spin system, as we shall discuss shortly.

\section{In the spatial domain}
\label{sec:discussion}

Let us conclude by interpreting the results derived above in the spatial domain,
where time is now considered as a spatial coordinate.
In this framework, the tied-down (or pinning) condition is very natural since it amounts to saying that the size of the finite system is fixed.

The $\sigma_{T}$ process defined at the beginning of section~\ref{sec:2time} 
now represents a one-dimensional spin system consisting of
a fluctuating number $N_L$ of spin domains spanning the total size of the system, denoted by $L$ in this context.
These domains have lengths $\tau_i$, which are discrete random variables with a common distribution denoted by $f_{\ell}=\prob(\tau=\ell)$.
The probability associated to the realisation $\{\ell_1,\dots,\ell_n\}$ of the sequence of $N_L=n$ intervals $\{\tau_1,\dots,\tau_n\}$, is given by the transcription in the spatial domain of~(\ref{eq:fstar})
\beq\label{eq:fstarspin}
p(L,\ell_1,\dots,\ell_n,n)
=\frac{f_{\ell_1}\dots f_{\ell_n}\delta\left(\sum_{i=1}^{n}\ell_i,L\right)}{Z(L)},
\eeq
where the Kronecker delta $\delta(i,j)=1$ if $i=j$ and $0$ otherwise.
The denominator is\footnote{For the tied-down random walk, starting and ending at the origin,~(\ref{eq:fstarspin}) and (\ref{eq:ZL}) have simple interpretations.
The former is the joint probability of a configuration for a walk of $L$ steps, the latter is the probability of return of the walk at time $L$ (where $L$ is necessarily even)~\cite{cgwendel}.}
\beqa\label{eq:ZL}
Z(L)
&=&\sum_{n\ge0}\sum_{\ell_1\dots\ell_n}f_{\ell_1}\dots f_{\ell_n}\delta\Big(\sum_{i=1}^n\ell_i,L\Big).
\eeqa

Equation~(\ref{eq:fstarspin}) can be interpreted as the Boltzmann distribution of an equilibrium model with Hamiltonian (or energy)
\beq\label{eq:Eequilib}
E(n,\{\ell_i\})=-\frac{1}{\beta}\sum_{i=1}^n\ln f_{\ell_i},
\eeq
where $(n,\{\ell_i\})$ is a realisation of the set of observables (number of domains, lengths of domains), with the constraint that the lengths of domains sum up to $L$.
In this context, $Z(L)$ is simply the partition function.
The expression~(\ref{eq:Eequilib}) is precisely the energy, at criticality, of the model defined in~\cite{bar1,bar2}, with the specific choice
\beq
f_{\ell}=\frac{1}{\zeta(c)}\frac{1}{\ell^c},
\eeq
where $c$ plays the role of $1+\theta$ and $\zeta(c)$ is the Riemann zeta function $\zeta(c)=\sum_{\ell\ge1} \ell^{-c}$.
This model is itself a simplified version of the Poland-Scheraga model~\cite{poland}, where the bubbles are seen as spin domains.

\begin{figure}[h]
\begin{center}
\includegraphics[angle=0,width=1\linewidth]{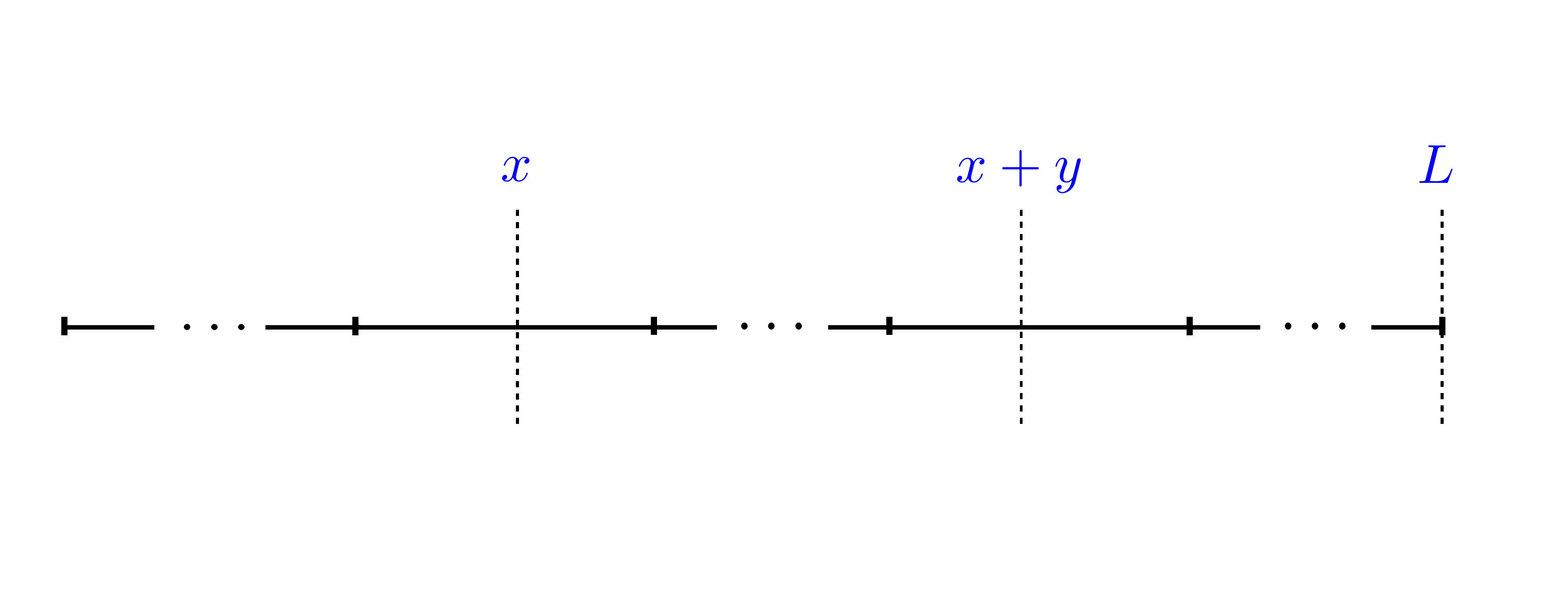}
\caption{Spatial coordinates defining the pair correlation function $\langle\sigma_x\sigma_{x+y}\rangle$.
\label{fig:corrSpin}}
\end{center}
\end{figure}

The two-time correlation $C(t,T,T+T')=\langle\sigma_{T}\sigma_{T+T'}\rangle$ becomes the spatial pair correlation function 
$C(L,x,x+y)=\langle\sigma_{x}\sigma_{x+y}\rangle$ (see figure~\ref{fig:corrSpin}).
The scaled quantity $M_t$ has the interpretation of the magnetisation of the spin system, i.e., (skipping details about the value of the spin located at the origin),
\beq
M_L=\frac{1}{L}\sum_{i=1}^{N_L}(-)^{i}\ell_i.
\eeq
The transcription of~(\ref{eq:fdpo}) yields the pair correlation function in the Porod regime $y\ll x\sim L$ ($L$ is large) 
\beqa\label{eq:porod}
\langle\sigma_x\sigma_{x+y}\rangle
&\approx &
1-\frac{\sin\pi\theta}{\pi(1-\theta)}\left[\frac{x}{L}\left(1-\frac{x}{L}\right)\right]^{\theta-1}
\left(\frac{y}{L}\right)^{1-\theta}
\nonumber\\
&\approx&1-\frac{\sin\pi\theta}{\pi(1-\theta)}\left(\frac{yL}{x(L-x)}\right)^{1-\theta}.
\eeqa
This expression is universal.
Likewise, the probability for two spins at distance $y$ apart to belong to the same domain is given in the persistence regime $x\ll y\sim L$ by the transcription of~(\ref{eq:fdpoP}), namely
\beqa
p_0(L,x,x+y)
&\approx&\frac{\sin\pi\theta}{\pi\theta}
\left(\frac{yL}{x(L-y)}\right)^{-\theta},
\eeqa
which is also universal.

The transcription of the results of section~\ref{sec:occup} predicts that the critical magnetisation is fluctuating in the thermodynamical limit, with a broad distribution $f_M(m)$ given by~(\ref{eq:fM}) and~(\ref{eq:fX}) (see figure~\ref{fig:occ}), whenever the tail exponent $\theta$ of the distribution of domain sizes $f_{\ell}$ is less than one (or $1<c<2$ for the exponent $c$).
The distribution $f_M(m)$ is universal, i.e., does not depend on the details of $f_{\ell}$.
We refer to~\cite{cgwendel} for a study of the distribution of the number of domains $N_L$.

The results above also provide some answers to issues raised in the past in the field of stochastically evolving surfaces.
In~\cite{das1,das2} coarse-grained depth models for Edwards-Wilkinson and KPZ surfaces are considered.
For one of them (the CD2 model in the classification of~\cite{das1,das2}) the surface profile is related to the tied-down random walk (corresponding to $\theta=1/2$).
The expression~(\ref{eq:porod}) can therefore be interpreted as the pair correlation function of this model in the Porod regime.
The prediction $\langle\sigma_x\sigma_{x+y}\rangle-1\sim (y/L)^{1-\theta}$ given in~\cite{das1,das2} for $0<\theta<1$ should also be compared to~(\ref{eq:porod}).
The largest interval for the CD2 model is found in~\cite{das1,das2}  by numerical simulations to satisfy $\tau_{\max}^{(1)}/L\approx 0.48 $, the second largest to satisfy 
$\tau_{\max}^{(2)}/L\approx 0.16 $.
These values are consistent with the analytical predictions $\tau_{\max}^{(1)}/L\approx 0.483498\ldots $, $\tau_{\max}^{(2)}/L\approx 0.159987\ldots$ obtained from~\cite{cgwendel,veto} (see also~\cite{bar}).
Finally, the existence of a broad distribution for the magnetisation of the tied-down renewal process (see~(\ref{eq:fM}) and~(\ref{eq:fX})), which can be seen as a generalisation of the CD2 model with a varying exponent $\theta$, is in line with the expected phenomenology put forward in~\cite{das1,das2,barma1,barma2} for fluctuation-dominated phase ordering phenomena.
Thus tied-down renewal processes with power-law distribution of intervals are minimal processes implementing a number of the expected characteristics of fluctuation-dominated phase ordering phenomena.

\ack
I am grateful to M Barma, J M Luck and D Mukamel for interesting discussions.

\appendix

\section{Derivation of equation~(\ref{eq:fElap})}
\label{app:fElap}

The number $N_{T}$ of events up to time $T$ takes the values $m=0,1,\dots$ and the number $N_t$ of events up to time $t$ takes the values $n=m+1,m+2,\dots$ (see figure~\ref{fig:figEA}).
Consider the probability density
\beqa
\fl f_{E,N_T,N_t}(t,T,m,n,y)
&=&\frac{{{\rm d}}}{{{\rm d}} y}\prob(E_{T}<y,N_T=m,N_t=n|t_{N_t}=t)
\nonumber\\
&=&\langle\delta(y-t_{m+1}+T)I(t_m<T<t_{m+1})\rangle,
\label{eq:fENN}
\eeqa
where $I(\cdot)$ is the indicator function of the event inside the parentheses.
Then by summation upon $m$ and $n$ we get the density of $E_T$
\beq
f_{E}(t,T,y)=\sum_{m\ge0}\sum_{n\ge m+1}f_{E,N_T,N_t}(t,T,m,n,y).
\eeq
The Laplace transform of~(\ref{eq:fENN}) with respect to $t,T,y$ (with $s,u,v$ conjugate to these variables) reads
\beq
\fl\lap{t,T,y}f_{E,N_T,N_t}(t,T,m,n,y)
=\lap{t}\langle\int_{t_m}^{t_{m+1}}{{\rm d}} T\,\e^{-uT}\e^{-v(t_{m+1}-T)}\rangle.
\eeq
Its numerator is
\beqa
\fl\lap{t,T,y}f_{E,N_T,N_t}(t,T,m,n,y)\num
&=&\int_0^\infty\left(\prod_{i=1}^n{\rm d}\ell_i\rho(\ell_i)\e^{-s\ell_i}\right)\e^{-vt_{m+1}}\int_{t_m}^{t_{m+1}}{{\rm d}} T\,\e^{-T(u-v)}
\nonumber\\
&=&\hat\rho(s+u)^m\hat\rho(s)^{n-m-1}\frac{\hat\rho(s+u)-\hat\rho(s+v)}{v-u}.
\label{eq:fENNlap}
\eeqa
Summing upon $m$ and $n$ yields~(\ref{eq:fElap}).

Setting $v=0$ in~(\ref{eq:fENNlap}) yields the Laplace transform 
\beq
\fl\lap{t,T}\prob(N_T=m,N_t=n|t_{N_t}=t)\num=
\hat\rho(s+u)^m\hat\rho(s)^{n-m-1}\frac{\hat\rho(s)-\hat\rho(s+u)}{u}.
\eeq
Summing this expression upon $m$ and $n$ gives back~(\ref{eq:PNN}).

\section{Derivation of equation~(\ref{eq:fAlap})}
\label{app:fAlap}

This derivation is very similar to that given above for $f_E$.
Consider the probability density
\beqa
\fl f_{A,N_T,N_t}(t,T,m,n,y)
&=&\frac{{{\rm d}}}{{{\rm d}} y}\prob(A_{T}<y,N_T=m,N_t=n|t_{N_t}=t)
\nonumber\\
&=&\langle\delta(y-T+t_{m})I(t_m<T<t_{m+1})\rangle.
\label{eq:fANN}
\eeqa
The Laplace transform of~(\ref{eq:fANN}) with respect to $t,T,y$ (with $s,u,v$ conjugate to these variables) reads
\beq
\fl\lap{t,T,y}f_{A,N_T,N_t}(t,T,m,n,y)
=\lap{t}\langle\int_{t_m}^{t_{m+1}}{{\rm d}} T\,\e^{-uT}\e^{-v(T-t_{m})}\rangle.
\eeq
Its numerator is
\beqa
\fl\lap{t,T,y}f_{A,N_T,N_t}(t,T,m,n,y)\num
&=&\hat\rho(s+u)^m\hat\rho(s)^{n-m-1}\frac{\hat\rho(s)-\hat\rho(s+u+v)}{u+v}.
\label{eq:fANNlap}
\eeqa
Summing upon $m$ and $n$ yields~(\ref{eq:fAlap}).

\section{Derivations of equations~(\ref{eq:pm}) and (\ref{eq:p0lap})}
\label{app:pmtTT}

Let the number of events $N_{T}$ up to time $T$ take the value $m_1$, and the number of events $N(T,T+T')=N_{T+T'}-N_{T}$ between $T$ and $T+T'$ take the value $m_2$.
We consider the probability of this event (see figure~\ref{fig:2timeB})
\beqa
\prob(N_T=m_1,N(T,T+T')=m_2,N_t=n|t_{N_t}=t)
\nonumber\\
\fl=\langle I(t_{m_1}<T<t_{m_1+1}) I(t_{m_1+m_2}<T+T'<t_{m_1+m_2+1})\rangle.
\eeqa
Consider first the case $m_2\ge1$.
In Laplace space, where $s,u,v$ are conjugate to the temporal variables $t,T,T'$, a simple computation gives
\beqa
\lap{t,T,T'}\prob(N_T=m_1,N(T,T+T')=m_2,N_t=n|t_{N_t}=t)\num
\nonumber\\
\fl=\hat\rho(s)^{n-m_1-m_2-1}\hat\rho(s+u)^{m_1}\frac{\hat\rho(s+v)-\hat\rho(s+u)}{u-v}
\hat\rho(s+v)^{m_2-1}\frac{\hat\rho(s)-\hat\rho(s+v)}{v}.
\eeqa
Summing on $m_1$ from $0$ and on $n$ from $m_1+m_2+1$ yields
\beqa
\lap{t,T,T'}\prob(N(T,T+T')=m_2|t_{N_t}=t)\num
\nonumber\\
\fl=\frac{1}{1-\hat\rho(s)}\frac{1}{1-\hat\rho(s+u)}
\frac{\hat\rho(s+v)-\hat\rho(s+u)}{u-v}
\frac{\hat\rho(s)-\hat\rho(s+v)}{v}
\hat\rho(s+v)^{m_2-1},
\eeqa
which is~(\ref{eq:pm}) (with $m\equiv m_2$).

Consider now the case $m_2=0$.
We have likewise
\beqa
\prob(N_T=m_1,N(T,T+T')=0,N_t=n|t_{N_t}=t)
\nonumber\\
\fl=\langle I(t_{m_1}<T<t_{m_1+1}) I(T+T'<t_{m_1+1})\rangle.
\eeqa
In Laplace space, where $s,u,v$ are conjugate to the temporal variables $t,T,T'$, a simple computation gives
\beqa
\lap{t,T,T'}\prob(N_T=m_1,N(T,T+T')=0,N_t=n|t_{N_t}=t)\num
\nonumber\\
\fl=\hat\rho(s)^{n-m_1-1}\hat\rho(s+u)^{m_1}
\frac{1}{v}
\left[\frac{\hat\rho(s)-\hat\rho(s+u)}{u}
-\frac{\hat\rho(s+v)-\hat\rho(s+u)}{u-v}\right].
\eeqa
Summing on $m_1$ from $0$ and on $n$ from $m_1+1$ yields
\beqa
\lap{t,T,T'}\prob(N(T,T+T')=0|t_{N_t}=t)\num
\nonumber\\
\fl=\frac{1}{1-\hat\rho(s)}\frac{1}{1-\hat\rho(s+u)}
\frac{1}{v}
\left[\frac{\hat\rho(s)-\hat\rho(s+u)}{u}
-\frac{\hat\rho(s+v)-\hat\rho(s+u)}{u-v}\right],
\eeqa
which is~(\ref{eq:p0lap}).


\newpage

\section*{References}

\begin{thebibliography}{99}

\bibitem{wendel}
Wendel J G 1964 {\it Math. Scand.} {\bf 14} 21

\bibitem{cgwendel} Godr\`{e}che C 2017 {\it J. Phys. A} {\bf 50} 195003

\bibitem{gl2001} Godr\`{e}che C and Luck J M 2001 {\it J. Stat. Phys.} {\bf 104} 489

\bibitem{gms2015}
Godr{\`e}che C, Majumdar S N and Schehr G 2015 {\it J. Stat. Mech.} P03014

\bibitem{poland} Poland D and Scheraga H A 1966
{\it J. Chem. Phys.} {\bf 45} 1464

\bibitem{bar1} Bar A and Mukamel D 2014 {\it Phys. Rev. Lett.} {\bf 112} 015701

\bibitem{bar2} Bar A and Mukamel D 2014 {\it J. Stat. Mech.} P11001

\bibitem{das1} Das D and Barma M 2000 {\it Phys. Rev. Lett.} {\bf 85} 1602

\bibitem{das2} Das D, Barma M and Majumdar S N 2001 {\it Phys. Rev.~E} {\bf 64} 046126

\bibitem{cox} Cox D R 1962 \textit{Renewal theory} (London: Methuen)

\bibitem{feller} Feller W 1968 1971 \textit{An Introduction to Probability Theory
and its Applications} Volumes 1\&2 (New York: Wiley)

\bibitem{balda} Baldassari A, Bouchaud J P, Dornic I and Godr\`{e}che C 1999
{\it Phys. Rev.~E} \textbf{59} R20

\bibitem{bray} Bray A J 1994 {\it Adv. Phys.} {\bf 43} 357

\bibitem{yano1} Yano Y 2006 {\it Publ. RIMS Kyoto Univ.} {\bf 42} 787

\bibitem{yano} Yano K and Yano Y 2008 {\it Statist. Probab. Lett.} {\bf 78} 2175

\bibitem{veto} Szab\'o R and Vet\"o B 2016 {\it J. Stat. Phys.} {\bf 165} 1086

\bibitem{bar} Bar A, Majumdar S N, Schehr G and Mukamel D 2016 {\it Phys. Rev. E} {\bf 93} 052130

\bibitem{barma1} Barma M 2008 {\it Eur. Phys. J. B} {\bf 64} 387

\bibitem{barma2} Barma M private communication

\end{thebibliography}
\end{document}